\documentclass[11pt,twoside]{article}
\usepackage{asp2004}
\usepackage{psfig}
\usepackage{epsf}
\newcommand{\msun}{M$_{\odot}$}
\newcommand{\ha}{H$\alpha$}
\newcommand{\degree}{$^{\rm o}$}

\begin{document}
\title{Linear line polarimetry modelling of pre-main sequence stars}
\author{Jorick S. Vink$^1$, Janet E. Drew$^1$, Tim J. Harries$^2$, Rene D. Oudmaijer$^3$}
\affil{$^1$ Imperial College London, Physics Department, Prince Consort Road,\\
       ~~~ London SW7 2AZ, UK\\
       $^2$ University of Exeter, School of Physics, Stocker Road, Exeter\\ 
       ~~~  EX4 4QL, UK\\
       $^3$ University of Leeds, School of Physics \& Astronomy,\\
       ~~~  EC Stoner Building, Leeds LS2 9JT, UK}

\begin{abstract}
We present emission line polarimetry data and modelling relevant to the 
circumstellar geometry and kinematics around pre-main sequence stars. 
For a sample of both Herbig Ae/Be stars and T Tauri 
stars, we find that most show polarization changes across \ha,  
implying that flattened structures are common on the smallest 
scales -- and over a range of stellar masses. We also present Monte 
Carlo calculations of spectral line profiles scattered in 
rotating accretion disks. We consider both 
the case of a central star that emits line photons uniformly, as 
well as via hot spots. Intriguingly, the switch between a uniform point 
source and a finite-sized star results in a marked difference in 
the position angle variation across the line. 
Our models demonstrate the diagnostic potential of line polarimetry in 
determining the disk inclination and the size of the inner 
hole -- a spatial scale no other technique currently accesses.
\end{abstract}
\thispagestyle{plain}

\section{General Introduction}

It is believed that low-mass stars form through the collapse of 
an interstellar cloud.
During the subsequent pre-main sequence (PMS) T Tauri phase material is 
accreting from the disk onto the star, most likely 
through magnetospheric funnels (e.g. Johns-Krull et al.~1999).
Whilst this basic picture of star formation is relatively well
understood, problems relating to a star's angular momentum remain, 
as we have little information on the size of the disk inner hole.
For intermediate mass 
(2 -- 10 $M_\odot$) Herbig Ae/Be stars our knowledge becomes even more 
patchy, and for stars above 10 $M_\odot$ there 
is not even any consensus on the mode of star formation itself.

Traditionally, the switch between low-mass and high-mass star formation
has been thought to occur at the T~Tauri/Herbig boundary 
(at $\simeq$~2 \msun), since this is where low-mass T~Tauri stars possess convective envelopes, 
whilst intermediate mass Herbig Ae star envelopes are radiative. However, recent 
data indicate that such a division is no longer tenable: Herbig Ae stars
and T~Tauri stars have a range of characteristics in common, varying from the
presence of inverse P~Cygni profiles, indicative of active accretion (Catala et al. 1999), 
to the detections of line polarizations, signalling rotating accretion 
disks (Vink et al. 2002, 2003).   
It is clear that what is needed to understand star formation as a function of mass 
(and ultimately understand the IMF) is observations 
of the near-star environment over a wide range of young stellar objects.
Polarimetry across emission lines is just such a tool.

\begin{figure}[!ht]
\vspace{-0.5cm}
\plotfiddle{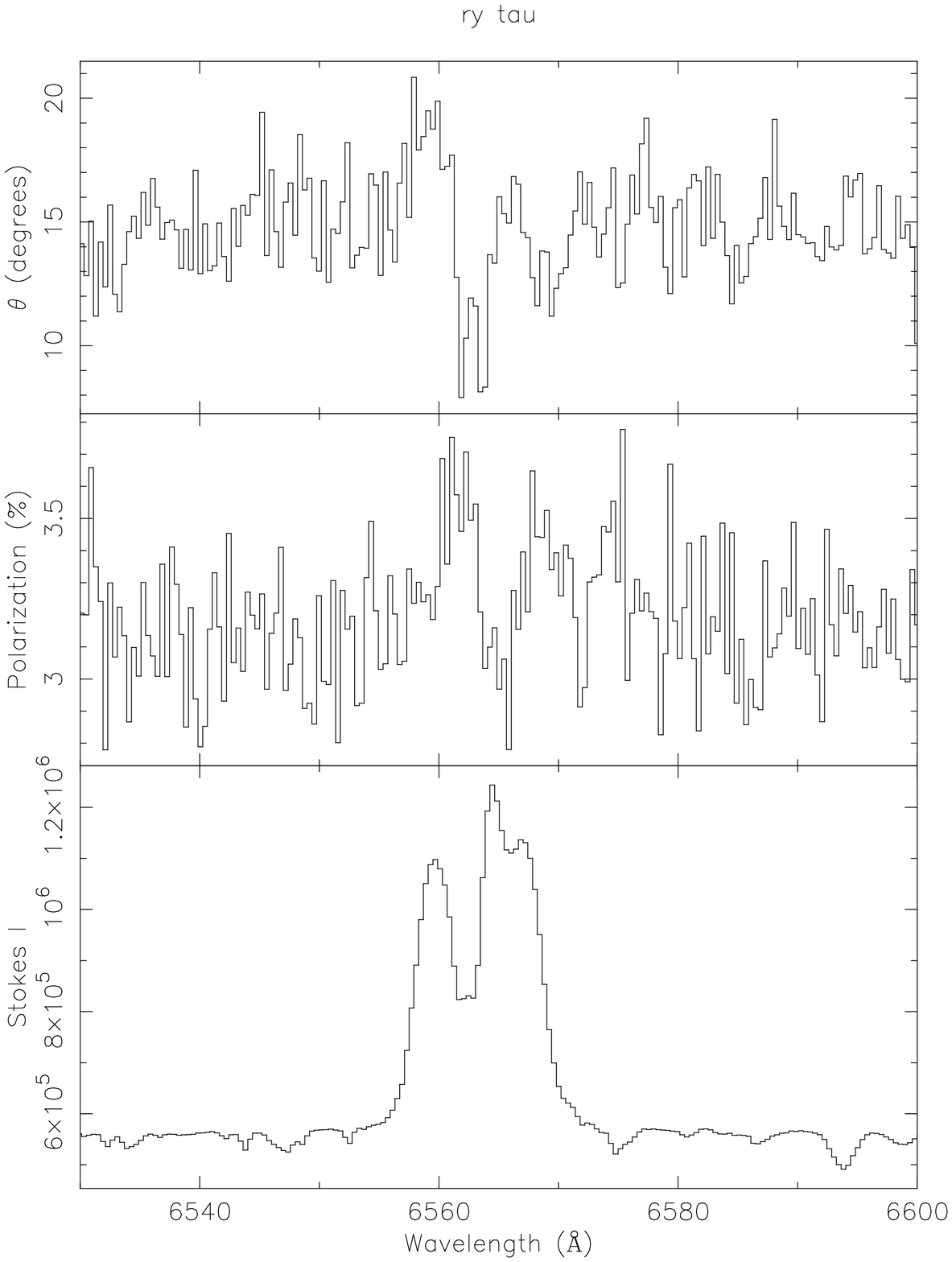}{0cm}{0}{30}{30}{-160pt}{-220pt}
\plotfiddle{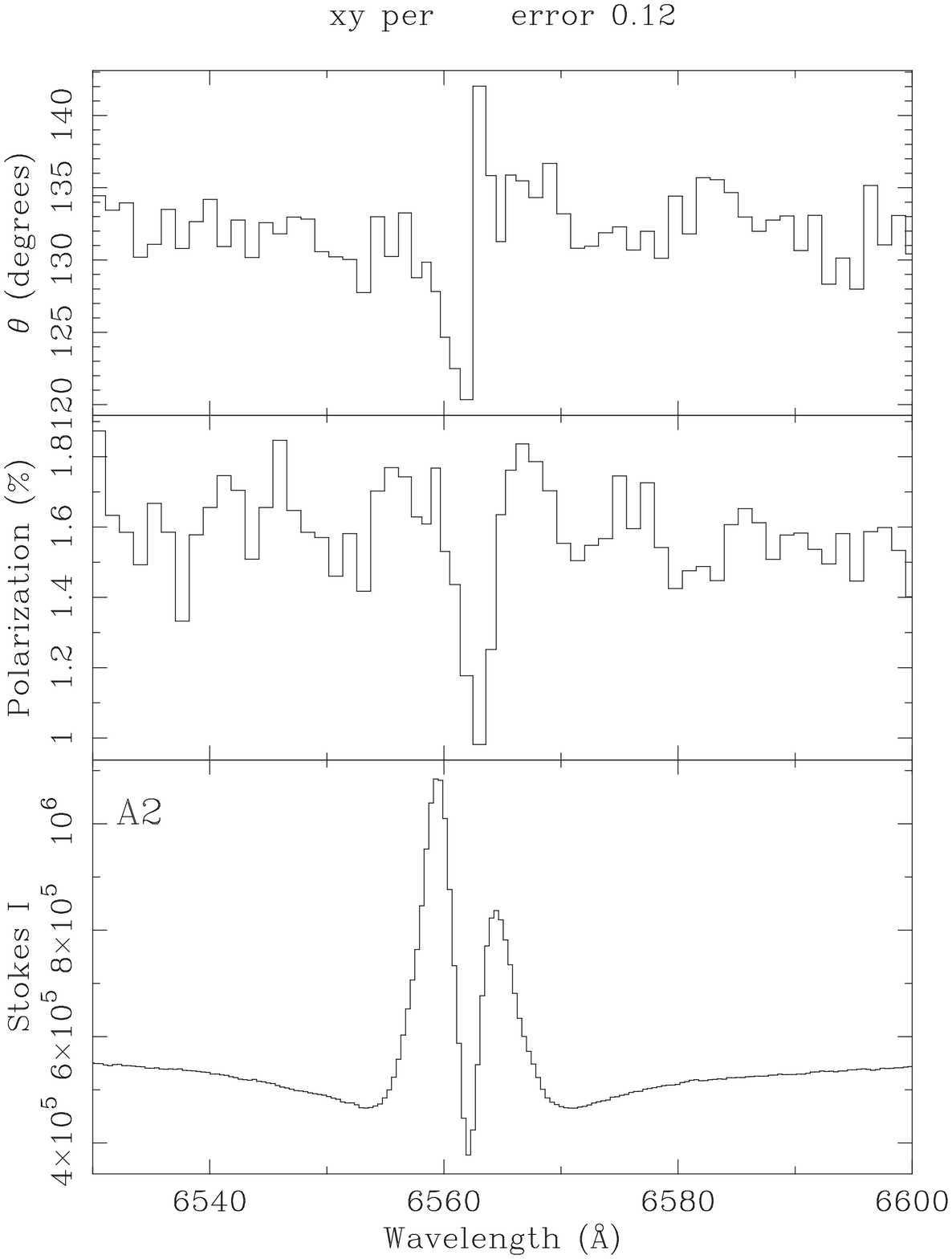}{0cm}{0}{30}{30}{0pt}{-195pt}
\plotfiddle{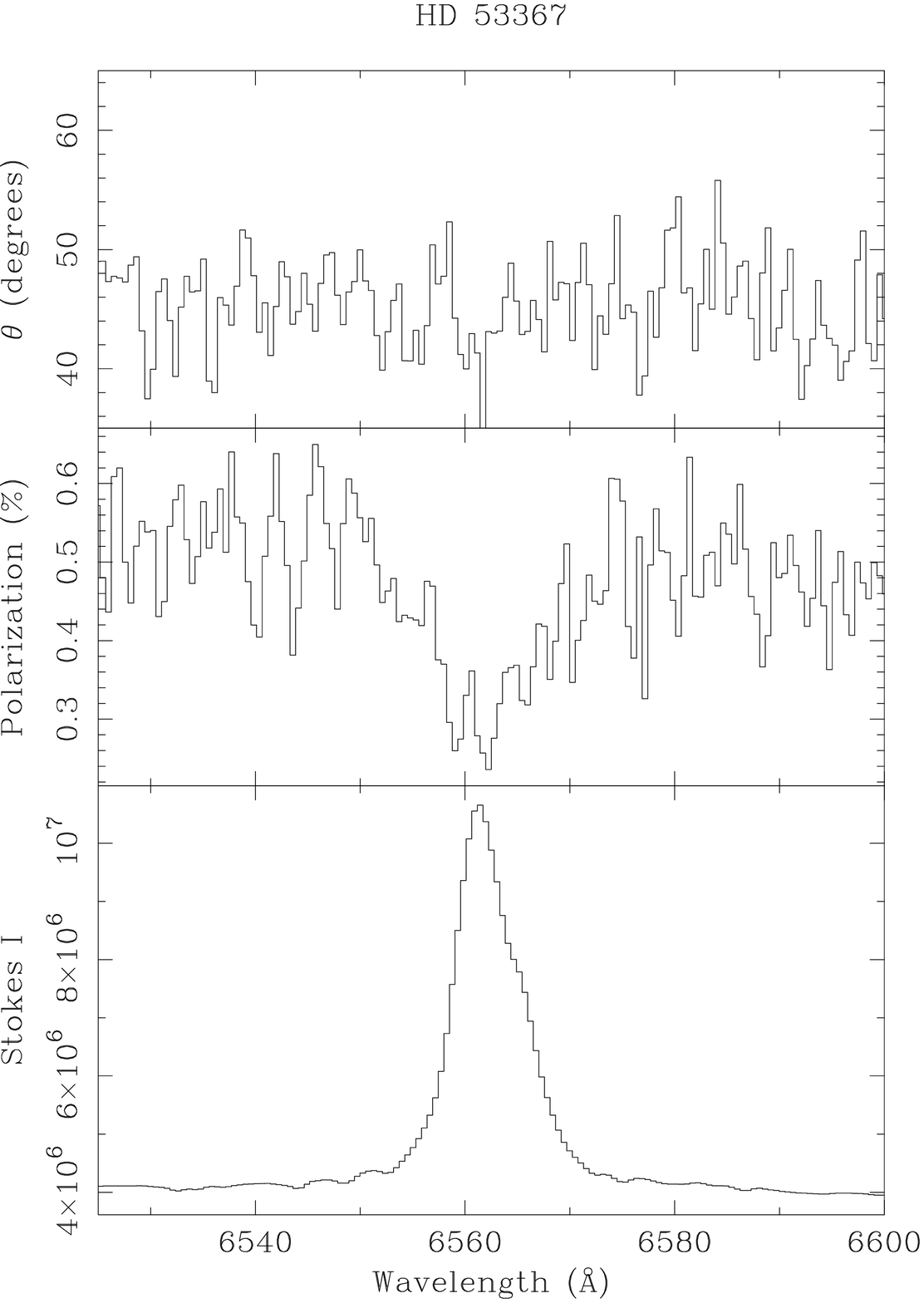}{7cm}{0}{30}{30}{-80pt}{-200pt}
\vspace{6.5cm}
\caption{Triplots of the observed polarization spectra of (a) the T Tauri star RY Tau
(Vink et al. 2003), the Herbig Ae star XY~Per (Vink et al. 2002), and the Herbig Be 
star HD~53367 (Oudmaijer \& Drew 1999). 
In all plots, the Stokes I spectrum is shown in the lowest panel,
the \%Pol is indicated in the middle panel, whilst the position
angle, $\theta$, is plotted in the upper panel.} 
\label{f_tri}
\end{figure}

\section{Line polarimetry}

Spectropolarimetry is a powerful tool to study the near-star regions 
of PMS stars and to determine their geometries. 
The technique has widely been applied to early-type stars, where circumstellar 
free electrons -- e.g. in a disk --  are able to polarize the continuum light 
more than the line photons. This is widely known as 'depolarization' (see Fig.~\ref{f_tri}(c) and 
Oudmaijer; these proceedings). 
In this case, the polarization angle (PA) of the polarization does 
not change across the line (if it is corrected for foreground), and the shape of the phenomenon in the $QU$ plane simply 
involves a straight line (independent of foreground).

In certain circumstances it is feasible that the {\it line} photons are scattered 
and polarized themselves (e.g. McLean 1979). 
Wood et al. (1993) performed an analytical study of the polarization and PA of 
a uniform point source that is scattered within a surrounding moving medium. Specifically, 
for a rotating disk, they found that the 
PA is no longer constant through the line, but rotates by a few degrees, resulting 
in a 'loop' in the $QU$ plane (see Fig.~\ref{f_qu} for examples), which they
attributed to stellar occultation. 
The diagnostic value of this PA flip ($QU$ loop) is that it is the direct 
signpost for the presence of rotation. 

\section{Data of T Tauri and Herbig Ae/Be stars}

\begin{figure}
\plottwo{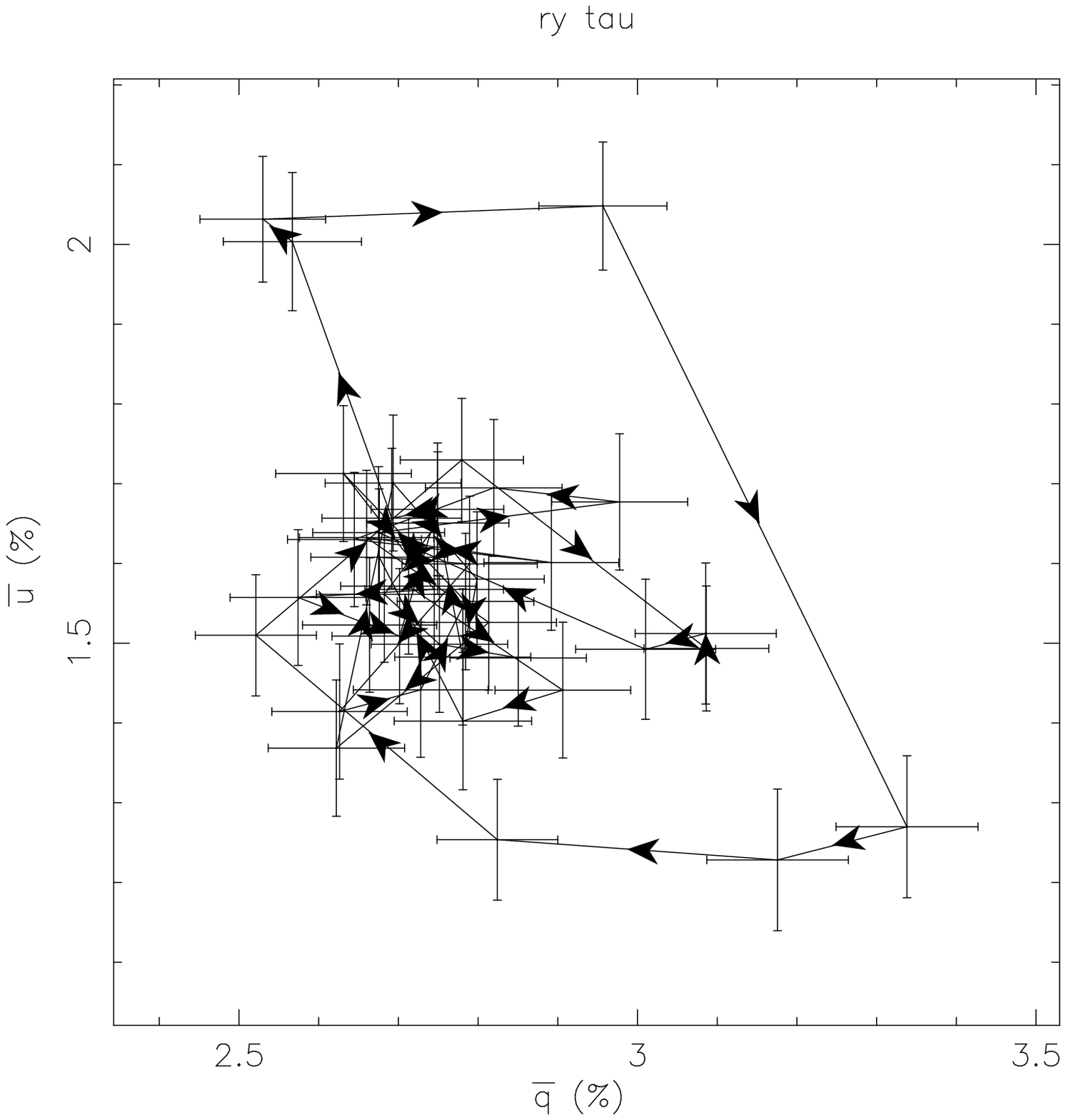}{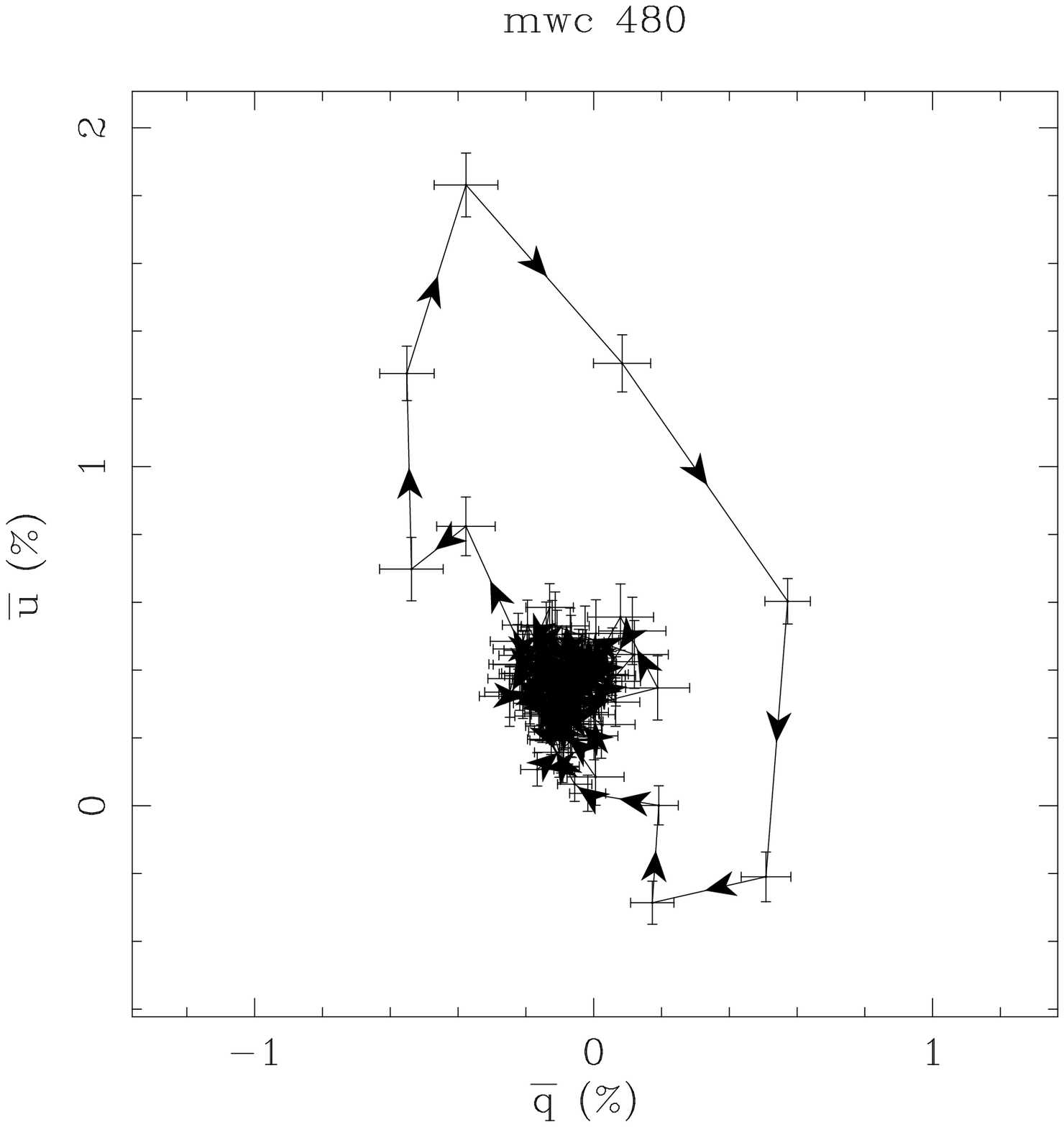}
\caption{$QU$ representations of the observed polarization spectra of the T~Tauri star 
RY~Tau and the Herbig Ae star MWC~480. The arrow denotes the sense of increasing wavelength.
Note the resemblance in the loops between the Herbig Ae and T Tauri star in 
the $QU$ diagram.}
\label{f_qu}
\end{figure}

In recent years, we have surveyed T~Tauri and Herbig Ae/Be stars 
spectropolarimetrically. For Herbig Be stars, the frequency of depolarizations was 
found to be essentially the same as was found for classical Be stars in the 1970s -- 
indicating they are embedded in electron scattering disks (see 
Fig.~\ref{f_tri}(c) for the Herbig Be star HD~53367
and Oudmaijer, these proceedings). 

When observing later spectral type PMS stars, one might perhaps 
expect to witness a sharp decrease in the number of polarization line effects,  
because the amount of free electrons is anticipated to drop. 
Furthermore, narrow-band filter work in the 1980s indicated a 
general absence of polarization changes across \ha\ in T Tauri stars (e.g. Bastien 1982).
However, this is not the case when observing with higher spectral resolution. 
Typical polarization spectra with a resolution of $R$ $\simeq$ 9000 of the T~Tauri 
star RY~Tau and the Herbig Ae star XY~Per are presented in Figs.~\ref{f_tri}(a) and (b).
The S-shaped PA flips in these data are indicative of {\it line} polarization, and 
the resulting loops in the $QU$ plane (Fig.~\ref{f_qu}) show that a rotating disk 
geometry is the dominant factor in producing the line effects in the majority of 
Herbig Ae (9/11; Vink et al. 2002) and T Tauri stars (9/10; Vink et al. 2004, 
in preparation).

 
\section{Line polarization models}

Given the common occurrence of these $QU$ loops in Herbig Ae and T~Tauri 
stars, we are developing polarization models of line emission 
scattered off rotating disks. In particular, we employ the 3D Monte Carlo model {\sc torus} 
(Harries 2000) to simulate both uniform as well as asymmetric illumination (by two diametrically 
opposed hot spots) onto a rotating scattering disk with, or without, a significant inner hole. 
These options of truncating the inner disk, and studying 
illumination via hot spots, are motivated by the growing interest in the 
magnetic accretion models for both T~Tauri (e.g. Edwards et al. 1994), and 
Herbig Ae stars (e.g. Vink et al. 2002). 

\begin{figure}
\plottwo{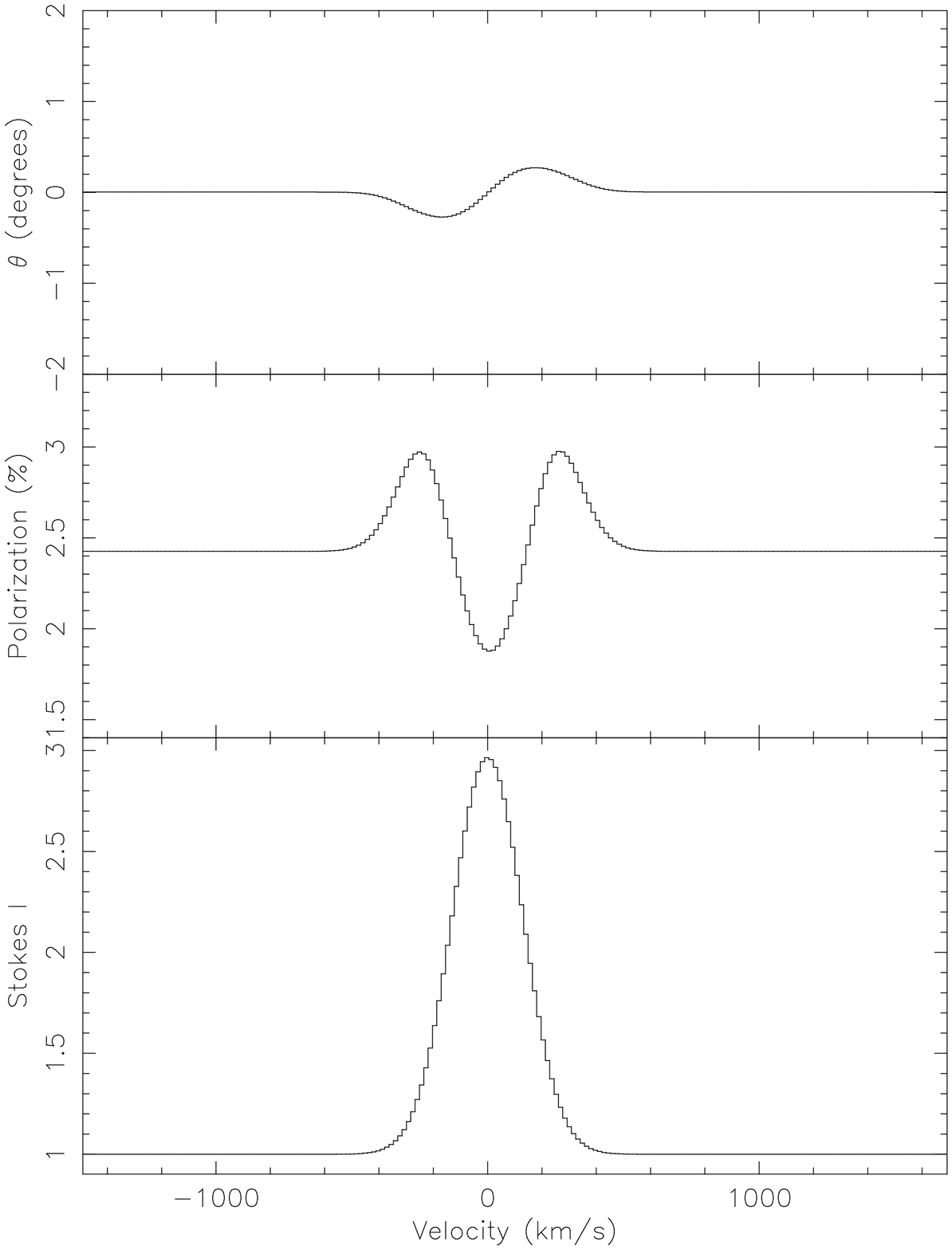}{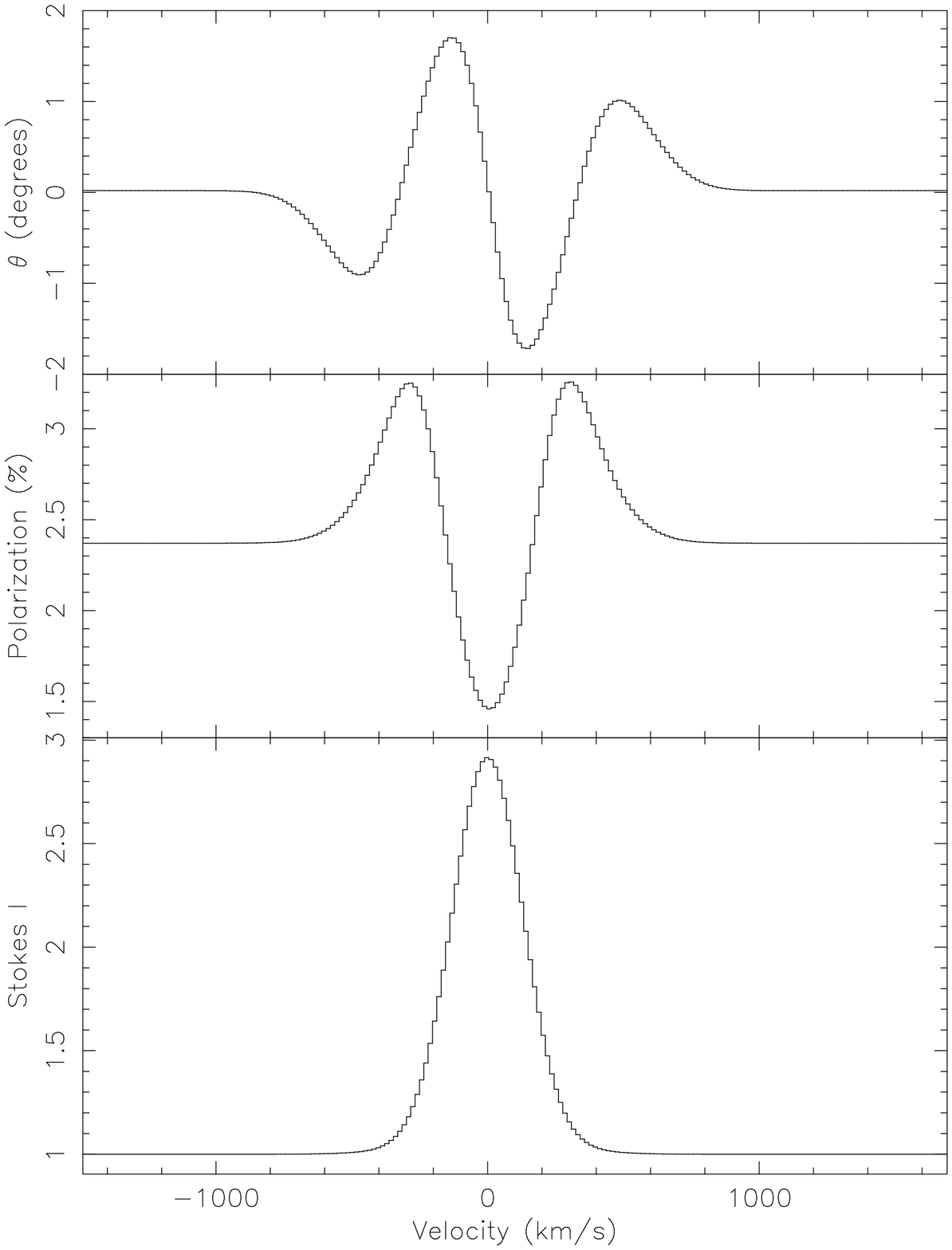}
\caption{Monte Carlo predictions of the line polarization for the cases of a finite-sized 
line-emitting star embedded in a scattering disk with (left) and without (right) an inner hole.
The disk is inclined at 45\degree\ and has an inner 
hole of 5 times the stellar radius (in the left panel).} 
\label{f_model}
\end{figure}

Figure~\ref{f_model} shows an intriguing result for the case of 
a uniformly radiating star. 
We find that there is a marked difference between scattering of line emission 
by a disk that reaches the stellar surface (Fig.~\ref{f_model}b), and a disk 
with a significant inner hole (Fig.~\ref{f_model}a). 
The single position-angle flip, seen on the left-hand side is similar to that 
predicted by the analytic models of Wood et al. (1993) -- but 
the double PA flip as seen on the right-hand side, 
associated with the undisrupted disk, is a surprise. 
This effect is due to 
the geometrically correct treatment of the finite-sized star interacting with 
the disk's rotational velocity field
(Vink et al. 2004).
Since a gradual increase of the hole size transforms the double rotations smoothly back 
into single ones -- as the line emission object approaches that of a point source -- our 
models demonstrate the diagnostic potential of line polarimetry in determining the disk 
inclination and the size of the inner hole.

By changing the configuration to a non-uniformly line emitting object, such 
as one where the emission originates from hot spots, 
we find that the line polarimetry depends strongly on the rotational 
phase. Stassun \& Wood (1999) have shown that the magnetic accretion model can account 
for the observed periodic changes in the PA and polarization of continuum light. 
However, photopolarimetry does not provide diagnostics of the disk truncation radius.
We have therefore extended Stassun \& Wood type-models to spectral lines. First 
results indicate that there are significant changes in the shapes and amplitudes 
of the PA and polarization across the spectral line as a function of rotational phase.  

\section{Summary}

We have presented data and modelling results of line polarimetry 
for PMS stars.
For the Herbig Ae/Be stars, we found that 
a large majority show a line effect -- indicating flattened 
circumstellar geometries.
Interestingly, we found a marked difference between the Herbig Be 
stars and groups of later spectral type. 
For the Herbig Be stars, electron scattering disks can explain the 
depolarisations. 
At lower masses, more complex behaviour appears across \ha. Here
the concept of compact line emission scattered off a rotating disk
may explain the observed $QU$ loops.

We have also presented polarimetric line profiles for scattering off accretion 
disks calculated with a Monte Carlo code. We considered the cases of a  
central object that emits line photons (a) uniformly, and (b) via hot 
spots only. For case (a), the switch between a point source and a finite-star 
photon source results in a surprising difference in the shapes of the 
predicted position angles. Most notably, we find double PA rotations. 
For case (b), emission from stellar hot spots, 
we find polarization signatures that are strongly dependent on 
hot spot phase with respect to the observer. 
Rotational modulation of line polarization shows great promise for 
unravelling the complexities occurring in the circumstellar environments   
around young stars.

\end{document}